\documentstyle[aps,prb]{revtex} 
\begin{document}
\title{Erratum: Kinetic theory of spin transport in n-type
  semiconductor quantum wells \ \ \ \ 
  [J. Appl. Phys. {\bf 93}, 410 (2003)]
}
\author{M. Q. Weng and M. W. Wu}

\bigskip

\maketitle
\bigskip\bigskip

There are some misprints in the text of our paper [JAP {\bf 93}, 410 (2003)]. 
They are corrected as following. 

The Poisson equation (5) in the paper should be
\setcounter{equation}{4}
\begin{equation}
  {{\bf \nabla}}^2_{{\bf r}}{\Psi}({\bf r},t) 
  = e[n({\bf r},t)-n_0({\bf r})]/\varepsilon_0,
\end{equation}
The screen constant $\kappa$ in the Eq. (8) should be
$\kappa^2=6\pi n_0({\bf r}) e^2/(a E_f)$. 

There are some misprints in  Eqs. (9)-(12), the corrected
equations are
\setcounter{equation}{8}
\begin{equation}
  \label{eq:f_coh}
 \left.{\partial f_{\sigma}({\bf R},{\bf k},t)\over \partial t}\right|_c=
  2\mbox{Im}[\bar{\varepsilon}_{\sigma-\sigma}({\bf R},{\bf k},t)
  \rho_{-\sigma\sigma}({\bf R},{\bf k},t)] 
\end{equation}
\begin{eqnarray}
\left.{\partial \rho_{\sigma-\sigma}({\bf R},{\bf k},t)\over \partial
  t}\right|_c 
&=&-i[\bar{\varepsilon}_{\sigma\sigma}({\bf R},{\bf k},t)-
\bar{\varepsilon}_{-\sigma-\sigma}({\bf R},{\bf k},t)]
\rho_{\sigma-\sigma}({\bf R},{\bf k},t) \nonumber\\
&&-i\bar{\varepsilon}_{\sigma-\sigma}({\bf R},{\bf k},t)
[f_{-\sigma}({\bf R},{\bf k},t)-f_{\sigma}({\bf R},{\bf k},t)]
\label{eq:rho_coh}
\end{eqnarray}
\begin{eqnarray}
  \left.{\partial f_{\sigma}({\bf R},{\bf k},t)\over \partial t}\right|_s
  &&=\biggl\{
  -2\pi\sum_{{\bf q}q_z\lambda}|g_{{\bf q}q_z\lambda}|^2
  \delta(\varepsilon_{\bf k}-
         \varepsilon_{{\bf k}-{\bf q}}-
         \Omega_{{\bf q}q_z\lambda})
  \Bigl[N_{{\bf q}q_z\lambda}
       \bigl(f_{\sigma}({\bf R},{\bf k},t)-
             f_{\sigma}({\bf R},{\bf k}-{\bf q},t)\bigr)
  \nonumber\\ &&
  +f_{\sigma}({\bf R},{\bf k},t) 
   \bigl(1-f_{\sigma}({\bf R},{\bf k}-{\bf q},t)\bigr)
  -\mathtt{Re}\bigl(
       \rho_{\sigma-\sigma}({\bf R},{\bf k},t)
        \rho^{\ast}_{\sigma-\sigma}({\bf R},{\bf k}-{\bf q},t)
    \bigr)\Bigr]
  \nonumber\\ &&
  -2\pi N_i\sum_{{\bf q}}U_{\bf q}^2
  \delta(\varepsilon_{{\bf k}}-\varepsilon_{{\bf k}-{\bf q}})
  \Bigl[f_{\sigma}({\bf R},{\bf k},t)
       \bigl(1-f_{\sigma}({\bf R},{\bf k}-{\bf q},t)\bigr)
  -\mathtt{Re}\bigl(\rho_{\sigma-\sigma}({\bf R},{\bf k},t)
  \nonumber\\&&
  \times \rho^{\ast}_{\sigma-\sigma}({\bf R},{\bf k}-{\bf q},t)\bigr)
  \Bigr]
  -2\pi\sum_{{\bf k}^{\prime}{\bf q}\sigma^{\prime}}V_{{\bf q}}^2
  \delta(\varepsilon_{{\bf k}-{\bf q}}
  -\varepsilon_{{\bf k}}+\varepsilon_{{\bf k}^{\prime}}
  -\varepsilon_{{\bf k}^{\prime}-{\bf q}})
  \nonumber\\ && 
  \times \Bigl[
  \bigl(1-f_{\sigma}({\bf R},{\bf k}-{\bf q},t)\bigr)
  f_{\sigma}({\bf R},{\bf k},t)
  \bigl(1-f_{\sigma^{\prime}}({\bf R},{\bf k}^{\prime},t)\bigr)
  f_{\sigma^{\prime}}({\bf R},{\bf k}^{\prime}-{\bf q},t)
  \nonumber\\ &&
  +{1\over 2}\mathtt{Re}
  \bigl(\rho_{\sigma-\sigma}({\bf R},{\bf k}-{\bf q},t)
        \rho_{-\sigma\sigma}({\bf R},{\bf k},t)\bigr)
  \bigl(f_{\sigma^{\prime}}({\bf R},{\bf k}^{\prime},t)-
  f_{\sigma^{\prime}}({\bf R},{\bf k}^{\prime}-{\bf q},t)\bigr)
  \nonumber\\ &&
  +{1\over 2}\mathtt{Re}\bigl(
 \rho_{\sigma^{\prime}-\sigma^{\prime}}({\bf R},{\bf k}^{\prime},t)
 \rho_{-\sigma^{\prime}\sigma^{\prime}}({\bf R},
 {\bf k}^{\prime}-{\bf q},t)\bigr)
 \bigl(f_{\sigma}({\bf R},{\bf k},t)-
 f_{\sigma}({\bf R},{\bf k}-{\bf q},t)\bigr)\Bigr]\biggr\}\nonumber\\
&-&\biggl\{{\bf k}\leftrightarrow{\bf k}-{\bf q},
    {\bf k}^{\prime}\leftrightarrow{\bf k}^{\prime}-{\bf q}\biggr\}
  \label{eq:f_scatt}
\end{eqnarray}
and
\begin{eqnarray}
  &&\left.{\partial\rho_{\sigma-\sigma}({\bf R},{\bf k},t)\over
      \partial t}\right|_s\nonumber\\ 
  & &=\Biggl\{-\pi\sum_{{\bf q}q_z\lambda}g^2_{{\bf q}q_z\lambda}
  \delta(\varepsilon_{\bf k}-\varepsilon_{{\bf k}-{\bf q}}
  -\Omega_{{\bf q}q_z\lambda})
  \Bigl[
  \bigl(f_{\sigma}({\bf R},{\bf k},t)+
  f_{-\sigma}({\bf R},{\bf k},t)\bigr)
  \rho_{\sigma-\sigma}({\bf R},{\bf k}-{\bf q},t)\nonumber\\
  \!\!\!\!\! &&+\bigl(f_{\sigma}({\bf R},{\bf k}-{\bf q},t)
  +f_{-\sigma}({\bf R},{\bf k}-{\bf q},t)-2\bigr)
  \rho_{\sigma-\sigma}({\bf R},{\bf k},t)
  -2N_{{\bf q}q_z\lambda}
  \bigl(\rho_{\sigma-\sigma}({\bf R},{\bf k},t)
  -\rho_{\sigma-\sigma}({\bf R},{\bf k}-{\bf q},t)\bigr)
  \Bigr]\nonumber\\
  \!\!\!\!\!&&-\pi N_i\sum_{{\bf q}\lambda}
  U_{\bf q}^2
  \delta(\varepsilon_{{\bf k}}-\varepsilon_{{\bf k}-{\bf q}})
  \Bigl[\bigl(f_{\sigma}({\bf R},{\bf k},t)
  +f_{-\sigma}({\bf R},{\bf k},t)\bigr)
  \rho_{\sigma-\sigma}({\bf R},{\bf k}-{\bf q},t)
  \nonumber\\ &&
  \mbox{}
  -\bigl(2-f_{\sigma}({\bf R},{\bf k}-{\bf q},t) -
  f_{-\sigma}({\bf R},{\bf k}-{\bf q},t)\bigr)
  \rho_{\sigma-\sigma}({\bf R},{\bf k},t)
  \Bigr]
  \nonumber\\ &&
  -\pi\sum_{{\bf k}^{\prime}{\bf q}\sigma^{\prime}}V_{\bf q}^2
  \pi \delta(\varepsilon_{{\bf k}-{\bf q}}-\varepsilon_{{\bf k}}
            +\varepsilon_{{\bf k}^{\prime}}-
             \varepsilon_{{\bf k}^{\prime}-{\bf q}})
  \biggl\{
  \bigl[f_{\sigma}({\bf R},{\bf k}-{\bf q},t)
  \rho_{\sigma-\sigma}({\bf R},{\bf k},t)
  +\rho_{\sigma-\sigma}({\bf R},{\bf k}-{\bf q},t)
  f_{-\sigma}({\bf R},{\bf k},t) \bigr]
  \nonumber\\ &&
  \times \bigl[f_{\sigma^{\prime}}({\bf R},{\bf k}^{\prime},t)
  -f_{\sigma^{\prime}}({\bf R},{\bf k}^{\prime}-{\bf q},t)\bigr]
  +\rho_{\sigma-\sigma}({\bf R},{\bf k},t)\Bigl[
  \bigl(1-f_{\sigma^{\prime}}({\bf R},{\bf k}^{\prime},t)\bigr)
  f_{\sigma^{\prime}}({\bf R},{\bf k}^{\prime}-{\bf q},t)
  \nonumber\\ && 
  -\rho_{\sigma^{\prime}-\sigma^{\prime}}({\bf R},{\bf k}^{\prime},t)
  \rho_{-\sigma^{\prime}\sigma^{\prime}}({\bf R},
  {\bf k}^{\prime}-{\bf q},t)\Bigr]
  -\rho_{\sigma-\sigma}({\bf R},{\bf k}-{\bf q},t)
  \Bigr[f_{\sigma^{\prime}}({\bf R},{\bf k}^{\prime},t)
  \bigl(1-f_{\sigma^{\prime}}({\bf R},{\bf k}^{\prime}-{\bf q},t)\bigr)
  \nonumber\\ &&  
  -\rho_{\sigma^{\prime}-\sigma^{\prime}}({\bf R},{\bf k}^{\prime},t)
  \rho_{-\sigma^{\prime}\sigma^{\prime}}({\bf R},
  {\bf k}^{\prime}-{\bf q},t)\Bigr]\biggr\}\Biggr\}
  \mbox{}-\biggl\{{\bf k}\leftrightarrow{\bf k}-{\bf q},
  {\bf k}^{\prime}\leftrightarrow{\bf k}^{\prime}-{\bf q}\biggr\}
  \label{eq:rho_scatt}
\end{eqnarray}
respectively.

{\em The numerical results in the paper are based on the correct formulas.}

\end{document}